\documentclass[aps,prl,twocolumn,superscriptaddress,showpacs]{revtex4}
\usepackage{graphicx}
\begin{document}

\preprint{APR 2004-XX}

\title{Coulomb Promotion of Spin-Dependent Tunnelling}
\author{L. Y. Gorelik}
\email{gorelik@fy.chalmers.se} \affiliation{Department of Applied
Physics, Chalmers University of Technology, SE-412 96
G\"{o}teborg, Sweden}
\author{S. I. Kulinich}
\affiliation{Department of Applied Physics, Chalmers University of
Technology, SE-412 96 G\"{o}teborg, Sweden}
\affiliation{B.~I.~Verkin Institute for Low Temperature Physics
and Engineering, 47 Lenin Avenue, 61103 Kharkov, Ukraine}
\author{R. I. Shekhter}
\affiliation{Department of Physics, G\"{o}teborg University,
SE-412 96 G\"{o}teborg, Sweden}
\author{M. Jonson}
\affiliation{Department of Physics, G\"{o}teborg University,
SE-412 96 G\"{o}teborg, Sweden}
\author{V.~M.~Vinokur}
\affiliation{Materials Science Division, Argonne National
Laboratory, 9700 South Cass Av, Argonne, Illinois 6043}

\date{\today}

\begin{abstract}
We study transport of spin-polarized electrons through a magnetic
single-electron transistor (SET) in the presence of an external
magnetic field. Assuming the SET to have a nanometer size central
island with a single electron level we find that the interplay on
the island between coherent spin-flip dynamics and Coulomb
interactions can make the Coulomb correlations promote rather than
suppress the current through the device. We find the criteria for
this new phenomenon --- Coulomb promotion of spin-dependent
tunnelling --- to occur.
\end{abstract}

\pacs{73.23.-b, 73.40Gk}
\maketitle

Strong Coulomb correlations have important consequences for
electronic transport on the nanometer length scale. Coulomb
blockade (CB) of single electron tunnelling \cite{Sh,Be} is, {\em
e.g.}, the fundamental physical phenomenon behind single electron
transistor (SET) devices \cite{Rev}. The electron spin comes into
play if the source and drain electrodes in the SET structure are
made of magnetic material, allowing for the electrons that carry
the current to be spin polarized \cite{MR,MR1}. Materials with
nearly 100~\% spin polarization are now under intensive study
\cite{12} and experiments have established that transport of spin
polarized electrons is sensitive to the relative orientation of
the magnetization of the source and drain leads. This opens up for
a spin-valve effect, where an external magnetic field can be used
to control the current. Switching the magnetization in one lead
with respect to the other is one way of achieving such a control
\cite{Sun}. Another approach --- to be pursued here --- is to flip
the spin of electrons carrying current from one magnetic lead to
another via a central non-magnetic island ("Coulomb dot") in a SET
structure, keeping the lead polarizations fixed.

A giant magneto-conductance effect was suggested recently for
magnetic "shuttle" devices, where spin-polarized electrons
localized on a movable quantum dot are mechanically transported
through a region, where the electronic spin dynamics is totally
controlled by an external magnetic field \cite{Gor}. This
suggestion brings into focus the very important question of the
role of the spin dynamics in resonant electron tunnelling, where
electrons can be trapped in a resonant state
--- {\em e.g.} on a Coulomb dot --- for quite a long time.
Since the resonant electron level of the dot may be doubly
occupied by electrons with different spin, Coulomb blockade of
single electron tunnelling should have an important influence on
the spin-dependent resonant charge transfer. Understanding how
these two effects combine is of general fundamental interest in
the context of magnetic nanostructures.

\begin{figure}[htb]
  \includegraphics[scale=0.35]{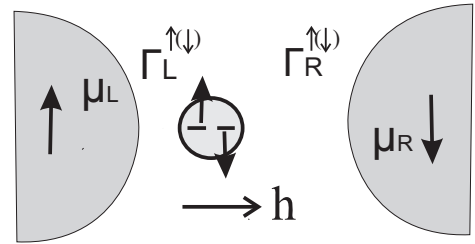}
  \vspace{-1 mm}
  \caption{ \label{fig:case12}Schematic view of the
nano-magnetic  SET device discussed in the text: a quantum dot,
with a single spin-degenerate electron level, is coupled to two
magnetic leads. The spin-dependent rates for tunnelling from the
dot to the left and right leads are
$\Gamma^{\uparrow(\downarrow)}_{L,R}$. The difference in
electrochemical potential between the leads $\mu_{L}-\mu_{R}= eV$
is caused by a bias voltage $V$. The dot is subject to an external
magnetic field $h$ oriented perpendicular to the direction of the
magnetization in the leads (arrows).}
   \vspace{-5 mm}
\end{figure}
 In this Letter will consider electronic transport through the
simple magnetic SET device shown in Fig.~1. Here a Coulomb dot,
subject to an external magnetic field, is located between two
spin-polarized leads. The external magnetic field is oriented
perpendicular to the polarizations in the leads and is taken to be
small enough not to affect the magnetization of the leads.
Assuming the dot is of nanometer size we consider only one
electron energy level on the dot. This level may, however,
accommodate two electrons of different spin orientation and the
strength of the Coulomb interaction between them significantly
affects the charge transfer through the device. The magnetic field
in its turn, by inducing coherent spin-flip dynamics on the dot,
actually promotes electronic transport if the magnetization in the
leads point in opposite directions. This effect is most
conspicuous when the leads are fully spin-polarized. In this case
an electron can tunnel from one lead to the other only if its spin
flips while the electron resides on the dot (without magnetic
field the current is completely blocked by a ``spin-blockade"). We
have found, that Coulomb correlations on the dot, if they are
strong enough to prevent a double occupancy of the resonant level,
significantly stimulates such spin-flip processes and hence
promote spin-dependent electronic tunnelling.

To understand this phenomenon qualitatively let us consider a
simplified set-up: a nonmagnetic quantum dot having a single
electronic level with energy $\epsilon$ is linked only to the
left, fully spin-polarized (up, say) metallic lead with
electrochemical potential $\mu_{L}$. Let $\mu_{L}-\epsilon$ be
much larger than the width $\Gamma_{L}$ of the dot level. Under
such conditions the spin-up state on the dot will be fully
occupied, while the spin-down state will be completely empty. If
now a magnetic field $h\ll\Gamma_{L}/\mu$ ($\mu$ is the Bohr
magneton) oriented perpendicular to the lead magnetization is
switched on at time $t=0$,
spin-flip processes that populate the spin-down state on the dot
will be induced. The characteristic time $\tau_{sf}$ for
populating the spin-down state is an important quantity. It turns
out that $\tau_{sf}$ strongly depends on whether or not Coulomb
interactions prevent a second electron from tunnelling onto the
dot during the spin-flip process.

To see this, let us first consider the Coulomb blockade regime
(CB-regime), where the energy difference $\mu_{L}-\epsilon$ is
smaller than the Coulomb interaction energy $U/2$ between two dot
electrons. The tunnelling of a second electron onto the dot is
then blocked and the population of the spin-down state is simply
controlled by the coherent spin dynamics of the one electron
already there. The probability amplitude $A_{sf}$ for a spin-flip
transition increases linearly with time, $A_{sf}=t/\tau_{h}$ (here
$\tau_{h}\equiv\hbar/ \mu h$), and the probability
$\rho_{\downarrow}$ to find the electron in the spin-down state is
\begin{equation}\label{1}
    \rho_{\downarrow}(t)=\vert A_{sf}\vert^2 = (t/\tau_{h})^{2}\,.
\end{equation}
The spin-flip time $\tau_{sf}$ may be estimated from the condition
$\rho_{\downarrow}(\tau_{sf})=1$. Hence in the CB-regime
$\tau_{sf}\equiv\tau_{c}\simeq \tau_{h}$.

If on the other hand $\mu_{L}-\epsilon > U$, we are in the ``free"
regime where the Coulomb blockade is lifted and a second (spin-up)
electron can tunnel onto the dot if there is a finite probability
for its spin-up state to be unoccupied. This process couples the
electronic state on the dot to a large number of states in the
lead and breaks the coherence of any ongoing evolution of the
on-dot spin state after a time $\Delta t\simeq\hbar/\Gamma_{L}$.
The probability for a spin flip to occur during this time is
$P_{sf}(\Delta t) \equiv |A_{sf}(t=\Delta t)|^{2}= (\Delta
t/\tau_{h})^{2}$. For longer times $t$ the probabilities for a
spin-flip to occur in $t/\Delta t$ coherent time intervals add
incoherently. Therefore, if the Coulomb blockade is lifted, the
probability to find the dot electron in the spin-down state can be
written  as
\begin{equation}\label{2}
    \rho_{\downarrow}(t)\simeq (t/\Delta t)P_{sf}(\Delta
    t) = (t\Delta t)/\tau_{h}^{2}\,,
\end{equation}
and the spin-flip time $\tau_{f}\simeq \tau_{c}\,(\tau_{h}/\Delta
t)$. It follows that $\tau_{c}\ll\tau_{f}$ in a weak magnetic
field ($h\ll\Gamma_{L}/\mu$). Hence, if the tunnelling of a second
electron is blocked, the probability for the spin of the electron
already on the dot to flip is strongly enhanced.

Next we extend our qualitative discussion to the tunnelling of
electrons through the entire SET device. For this purpose we
switch on the coupling between the spin-down dot state and states
in the right lead held at chemical potential $\mu_{R}$. If
$\mu_{R}\ll\epsilon - \Gamma_{R}$, where $\Gamma_{R}/\hbar$ is the
tunnelling rate between dot and right lead, the spin-down dot
electron can tunnel to an empty state in the right lead. The
resulting current through the SET is given by the product of the
probability to find the dot electron in the spin-down state and
the tunnelling rate $\Gamma_{R}/\hbar$. If the spin flip rate is
much smaller than $\Gamma_{R}/\hbar$ the population of the
spin-down state can be estimated as
$\rho_{\downarrow}(t=\hbar/\Gamma_{R})$. Furthermore, since
electron exchange with the right lead  also restricts the coherent
spin-flip time one has to put  $\Delta t\min\{\hbar/\Gamma_{L},\hbar/\Gamma_{R}\}$ in Eq.~(\ref{2}).
Therefore, in a strongly asymmetric situation, when the ratio
$\Gamma_{R}/\Gamma_{L}$ is very different from one, the current
through the SET is given by the expression
\begin{equation}\label{I}
    I = \frac{e}{\hbar}\frac{(\mu h)^{2}}{\Gamma_{R}}\left\{
\begin{array}{c}
 1 \qquad\qquad\quad\qquad\textmd{CB regime}\,\,(c) \\
 \min\{1,\Gamma_{R}/\Gamma_{L}\}\,\,\,\textmd{free regime}\,\,(f)\\
\end{array}
\right.
\end{equation}
One concludes that, when $\Gamma_{R}/\Gamma_{L}\ll 1$, the current
in the CB-regime is larger by a factor $\Gamma_{L}/\Gamma_{R}$
than in the ``free" regime, where the CB is lifted. We will refer
to this as ``Coulomb promotion" of spin-dependent tunnelling. From
Eq.~(\ref{I}) it also follows that the current-voltage curve is
strongly asymmetric in the CB-regime; the current changes by a
factor $\Gamma_{R}/\Gamma_{L}$ if the sign of the bias voltage is
reversed.

To provide a quantitative description of the Coulomb promotion
phenomenon discussed above we consider a SET structure with
magnetic leads polarized along the same $z$-direction, and a
central island (dot) with a single electron energy level subject
to an external magnetic field $\vec{h}$. The Hamiltonian
$\hat{\cal{H}}$ for our system is
\begin{eqnarray}\label{01}
&&\hat{\cal H} = \hat {\cal H}_l + \hat {\cal H}_{d}+\hat {\cal
H}_T \,,\nonumber
\\&&\hat {\cal H}_l= \sum_{\alpha, \sigma, \kappa}
\varepsilon_{\alpha, \sigma, \kappa} a^\dag_{\alpha,
\sigma,\kappa}  a_{\alpha,\sigma, \kappa} \,,\nonumber
\\&&\hat{\cal H}_d=\sum_\sigma \epsilon a^\dag_ \sigma
a_\sigma -\frac{U}{2} a^\dag_\uparrow  a^\dag_\downarrow
a_\uparrow a_\downarrow - \mu  \sum_{i,\sigma, \sigma'}h_{i}
a^\dag_{\sigma} \tau_{i}^{\sigma,\sigma'}a_{\sigma'} \nonumber
\\&& {\cal H}_{T}=\sum_{\alpha, \sigma, \kappa} T_\kappa(
a^\dag_{\alpha, \sigma,\kappa}  a_\sigma + \text{H.c.}) \,,
\end{eqnarray}
and has several terms. The first term describes noninteracting
electrons in the leads. Here $a^\dag_{\alpha, \sigma, \kappa}$
$(a_{\alpha, \sigma, \kappa})$ is the creation (annihilation)
operator for electrons in lead $\kappa$ with energy
$\varepsilon_{\alpha, \sigma, \kappa}$  and spin projection
$\sigma=(\uparrow,\downarrow)$. The electron density of states
$\nu^{\sigma}_{\kappa}$ in each lead is assumed to be independent
of energy but strongly dependent on spin direction. The electrons
in each lead are held at a constant electrochemical potential
$\mu_{L,R}=E_{F}\mp e V/2$, where $e$ is the charge of the
electron, $V>0$ is the bias voltage, and $E_{F}$ is the Fermi
energy of the ferromagnetic metal. The second term describes
electronic states in the dot, their coupling to the external
magnetic field $\vec{h}=(h_{x},0,h_{z})$ and intra-dot electron
correlations characterized by the Coulomb energy $U$; the operator
$a^\dag_\sigma$ $(a_\sigma)$ creates (destroys) an electron with
spin $\sigma$ and $\tau_{i}^{\sigma,\sigma'}$ are Pauli matrices
($i=x,y,z$). The last term represents spin-conserving tunnelling
of electrons between dot and leads.

We focus on the Coulomb promotion phenomenon discussed
qualitatively above and assume the Coulomb energy $U$ to be much
larger than both the thermal energy $kT$ and the width and Zeeman
splitting of the dot level. We furthermore take the energy level
on the dot to coincide with the Fermi energy of the leads,
$\epsilon = E_{F}$, and consider the current for bias voltages
$|V|\gg(\mu|\vec{h}|,\,\Gamma^{\sigma}_{\kappa},\,kT)$ and
$|V-U|\gg(\mu|\vec{h}|,\,\Gamma^{\sigma}_{\kappa},\,kT)$ (here
$\Gamma^{\sigma}_{\kappa}\equiv
2\pi\nu^{\sigma}_{\kappa}T^{2}_{\kappa}$ is the spin-dependent
level width associated with tunnelling to the lead $\kappa$). If
the external magnetic field is oriented along the polarization
axis, charge transfer between the leads conserve spin projection
and may be described within the ``orthodox" CB theory \cite{Sh}. A
magnetic field oriented perpendicular to the polarization axis, on
the other hand, generates coherent spin dynamics and makes the
spin degree of freedom relevant for the electron transport
problem.

The coupled processes of charge transfer and coherent spin
dynamics is governed by a quantum Master equation for the
corresponding reduced density operator $\hat{\rho}(t)$. It can be
derived from the Liouville-von Neumann equation for the total
system by projecting out the degrees of freedom associated with
the leads \cite{Naz,Gur,Dong,Suka}. The reduced density operator
$\hat{\rho}(t)$ obtained in this way acts on the Fock space of the
quantum dot which is spanned by the four basis vectors
$|0\rangle$, $|\uparrow\rangle \equiv a^\dag_{\uparrow}|0\rangle$,
$|\downarrow\rangle \equiv a^\dag_{\downarrow}|0\rangle$, and
$|2\rangle \equiv a^\dag_{\uparrow}a^\dag_{\downarrow}|0\rangle$.
In this basis the operator $\hat{\rho}(t)$ can be written as a
$4\times 4$ matrix. The diagonal elements $\rho_{0}=\langle
0|\hat{\rho}(t)|0\rangle$ and $\rho_{2}=\langle
2|\hat{\rho}(t)|2\rangle$ represent the probabilities for the dot
to be unoccupied and doubly occupied, respectively. The singly
occupied dot is described by the $2\times 2$ matrix block
$\hat{\rho}_{1}\equiv(\rho)_{\sigma,\sigma'}
\equiv\langle\sigma|\hat{\rho}|\sigma'\rangle$

For bias voltages satisfying the given conditions, the time
evolution of the probabilities $\rho_{0}$, $\rho_{2}$ and of the
density matrix $\hat{\rho}_{1}$ is determined by the system of
equations
\begin{eqnarray}\label{M-E}
 \dot{\rho}_{0}&=&-{\rm Tr}\{\hat{\Gamma}_{L}\}\rho_{0}+
{\rm Tr}\{\hat{\Gamma}_{R}\hat{\rho}_{1}\}\nonumber\\
\dot{\rho}_{2}&=&-{\rm
Tr}\{\hat{\Gamma}_{R}+\theta(U-V)\hat{\Gamma}_{L})\}\rho_{2}
+\theta(V-U){\rm Tr}\{\hat{\Gamma}_{L}\hat{\rho}_{1})\}\nonumber\\
 \dot{\hat{\rho}}_{1}&=&i\mu [h_{i}\hat{\tau}_{i},\hat{\rho}_{1}]
- \frac{1}{2}\{\hat{\Gamma}_{R},\hat{\rho}_{1}\}
-\frac{1}{2}\theta(V-U)\{\hat{\Gamma}_{L},\hat{\rho}_{1}\}\nonumber\\
&& +
\hat{\Gamma}_{L}\rho_{0}+(\hat{\Gamma}_{R}+\theta(U-V)\hat{\Gamma}_{L})\rho_{2}\,,
\end{eqnarray}
where $\hat{\Gamma}_{\kappa} (\Gamma_{\kappa}^{\uparrow}+\Gamma_{\kappa}^{\downarrow})\hat{I}/2+
(\Gamma_{\kappa}^{\uparrow}-\Gamma_{\kappa}^{\downarrow})\hat{\tau}_{z}/2$,
$\hat{\tau}_{i}$ are the Pauli matrixes, and we have set
$e=\hbar=1$.

The stationary solution of Eq.~(\ref{M-E}), and hence the
stationary current, significantly depends on the relation between
the bias voltage $V$ and the CB energy $U$. In particular, the
third equation
tells us that the CB significantly decreases the relaxation of the
non diagonal elements of $\hat{\rho}_{1}$. As these describe the
coherent spin flip dynamics, the spin-flip process turns out to be
faster in the CB-regime than in the ``free" regime without Coulomb
blockade.

Within our approximations the current $I$ through the SET can be
calculated from the formula
\begin{eqnarray}\label{cur}
I={\rm Tr}\{\hat{\Gamma}(\hat{\rho}_{1}+\rho_{2}\hat{I})\}\,.
\end{eqnarray}
It is convenient to think of the current as the sum of a
background current $I^{(0)}\equiv I(h=0)$ and a magnetic-field
promoted (MFP) current $J(h)\equiv I-I^{(0)}$. Substituting the
stationary solution of Eq.~(\ref{M-E}) into Eq.~(\ref{cur}), one
finds
\begin{equation}\label{Cur1}
I_{l}=I_{l}^{(0)}+J_{l}(h)=I_{l}^{(0)}+
J_{l}\frac{h_{x}^2}{h_{l}^2+h_{x}^2}\,,
\end{equation}
where the label $l=c,f$ indicates whether we are in the CB-regime
($V<U$) or in the "free" regime where the blockade has been lifted
($V>U$).  The quantities $I_{l}^{(0)}$, $J_{l}$ and $h_{l}$ do not
depend on the transverse magnetic field and are given by
\begin{eqnarray}\label{c}
 &&I_{c}^{(0)}=\frac{\Gamma_L\Gamma_R}{\Gamma_L+\Gamma}-J_{c},\,
J_{c}=\frac{\Gamma_L(\Gamma_L^\downarrow\Gamma_R^\uparrow-\Gamma_L^\uparrow
\Gamma_R^\downarrow)(\Gamma_R^\uparrow-\Gamma_R^\downarrow)}
{(\Gamma_L+\Gamma)(\Gamma^\uparrow\Gamma^\downarrow-
\Gamma_L^\downarrow\Gamma_L^\uparrow)}
\,,\nonumber\\
&&h_{c}^{2}=\mu^{-2}\frac{\Gamma_R(\Gamma^
\uparrow\Gamma^\downarrow- \Gamma_L^\downarrow\Gamma_L^\uparrow)}
{\Gamma_L+ \Gamma}\left[\frac{1}{4}+\frac{(\mu
h_{z})^2}{\Gamma_R^2} \right]\, ,\nonumber\\\
&&I_{f}^{(0)}=\frac{\Gamma_L\Gamma_R}{\Gamma}-J_{f},\,\,\,
J_{f}=\frac{(\Gamma_L^\uparrow\Gamma_R^\downarrow-
\Gamma_L^\downarrow\Gamma_R^\uparrow)^2}{\Gamma
\Gamma^\uparrow\Gamma^\downarrow}\,,\nonumber\\
&&h_{f}^{2}=\mu^{-2}\Gamma^\uparrow\Gamma^
\downarrow\left[\frac{1}{4}+\frac{(\mu h_{z})^2} {\Gamma^2}\right]
.
\end{eqnarray}
Here $\Gamma_\kappa=\sum_{\sigma} \Gamma_\kappa^\sigma$,
$\Gamma^{\sigma}=\sum_{\kappa} \Gamma_\kappa^\sigma$, and
$\Gamma=\sum_{\sigma,\kappa} \Gamma^\sigma_{\kappa}$.
\begin{figure}[htb]
  \label{Fig1}
  \includegraphics[scale=0.4]{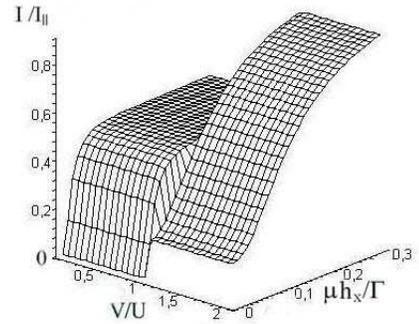}
  \vspace{-5 mm}
  \caption{ \label{fig:case22} Normalized current $I$
  through a
  magnetic SET structure as a function of transverse magnetic field $h$
  and bias voltage $V$. The polarizations of source and drain are antiparallel
  and described by $\beta = 0.98$, while
  the asymmetry parameter $\alpha = 0.9$. For a fixed
  magnetic field the current changes by a step at
  the CB threshold, where $V=U$.
  The background current at $h=0$ and the large-field current ($h\rightarrow\infty$)
  increase as expected when the CB is lifted.
  In contrast, Coulomb correlations promote spin-dependent tunnelling for intermediate
  fields; the current drops when the CB is lifted,
  which is a signature of the Coulomb promotion phenomenon described in the text.}
  \vspace{0 mm}
\end{figure}

Our results Eqs.~(\ref{Cur1}) and (\ref{c}) for the current are
valid for any values of $\vec{h}$ and $\Gamma_{\kappa}^{\sigma}$
that meet the conditions formulated above.  In this Letter we are
particularly interested in situations where the Coulomb promotion
phenomenon can be observed. Therefore, we assume in the analysis
to follow that the leads are made from the same magnetic material
and are polarized in opposite directions \cite{mis}. This implies
that $\Gamma_{L}^{\downarrow}/\Gamma_{L}^{\uparrow}\Gamma_{R}^{\uparrow}/\Gamma_{R}^{\downarrow}=(1-\beta)/(1+\beta)$,
where $\beta \subset [0,1]$ is a polarization parameter. The
limiting values $\beta = 1$ and $\beta=0$ denote fully polarized
and unpolarized leads, respectively. Furthermore, we take the
magnetic field to be directed perpendicular to the magnetization
in the leads.

From Eq.~(\ref{c}) it is then clear that
$J_{c}/h^{2}_{c}>J_{f}/h^{2}_{f}$ and $h_{f}>h_{c}$. Invoking in
addition Eq.~\ref{Cur1} we conclude that the magnetic-field
promoted (MFP) current is larger in the CB-regime than in the
"free" regime if $h<h_{c}$. In relatively small magnetic fields,
therefore, the MFP-current drops when the bias voltage lifts the
Coulomb blockade. This drop, shown in Fig.~2, is a signature of
the Coulomb promotion phenomenon.

The relative value of the drop $\Delta J(h) (J_{c}(h)-J_f(h))/I_{||}$ ($I_{||}\equiv I^{(0)}_{f}$) in the
MFP-current at $V=U$ depends on the magnetic field as well as on
the degree of spin polarization. It also depends on the asymmetry
of the SET structure, which we characterize by the parameter
$\alpha \equiv (\Gamma_{L}-\Gamma_{R})/\Gamma$. The  value $\Delta
J\equiv\max\Delta J(h)$ as a function of $\alpha$ and $\beta$ is
shown in Fig.~3. We note that the Coulomb promotion phenomenon is
most prominent for strongly asymmetric SET structures with highly
spin polarized leads, {\em i.e.} when $\alpha\simeq 1$ and $\beta
\simeq 1$. If the leads are fully polarized, $\beta =1$, the
background current vanishes. As a result, the MFP-current gives
the total current and the Coulomb promotion manifests itself as a
negative differential conductance at the bias voltage $V\approx
U$. It is interesting to note that if $\beta =1$, $\alpha
\simeq\pm 1$ and $h/h_{c}\ll 1$ (the case which was qualitatively
discussed above) then Eqs.~(\ref{Cur1}) and (\ref{c}) reproduce
Eq.~(\ref{I}) to leading order in small parameters. However, if
the leads are not completely spin polarized the Coulomb blockade
of the background current competes with the Coulomb  of the
MFP-current. As a result the negative differential conductance
appears only in a small region where the parameter $\beta > 0.9$
(see the insert in Fig.~3).
\begin{figure}[htb]
  \label{Fig2}
  \includegraphics[scale=0.38]{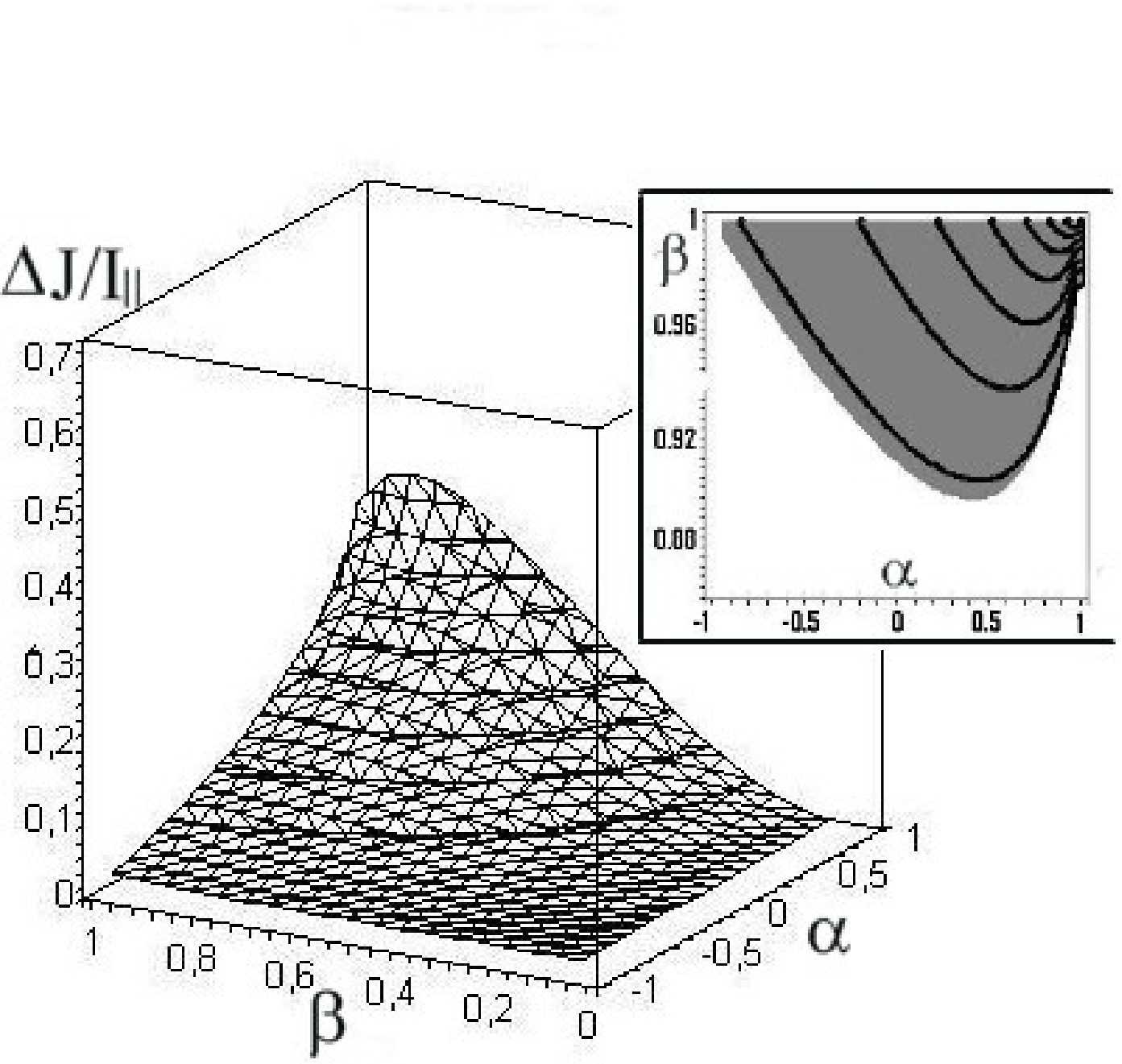}
  \vspace{+3 mm}
  \caption{ \label{fig:case12}
Drop of the magnetic-field promoted current $\Delta J$,
  caused by the lifting of the CB at $V=U$, as a function of the
  polarization and asymmetry parameters $\beta$ and $\alpha$ (see text).
  The domain in the $\alpha,\beta$-plane, where a negative differential
  conductance
  (NDC) may be observed (dark) is shown on the inset.
  The current drop associated with a NDC is of the order of $I_{||}$
  and peaks at $\beta=1,\,\alpha\simeq +1$.}
\vspace{-5 mm}
\end{figure}

In conclusion we have studied resonant tunnelling of
spin-polarized electrons through a magnetic SET device with a
central island subject to an external magnetic field directed
perpendicular to the magnetization in the leads. The combined
effects of spin-dependent tunnelling between a source and a drain
with antiparallel magnetizations and of Coulomb correlations were
considered. We find that a Coulomb blockade preventing the single
electron level on the central "Coulomb dot" to be doubly occupied
may significantly stimulate the transport of electrons through the
device. This effect gives rise to a new phenomenon -- Coulomb
promotion of spin-dependent tunnelling.

This work was supported in part by the European Commission through
project FP6-003673 CANEL of the IST Priority. The views expressed
in this publication are those of the authors and do not
necessarily reflect the official European Commission's view on the
subject. Financial support from the Swedish SSF, the Swedish VR,
and from the U.S. DOE Office of Science through contract No.
W-31-109-ENG-38 is also gratefully acknowledged.


\begin{thebibliography}{0}
\bibitem{mis} For
SET configurations with identical leads that are either
unpolarized or polarized in the same direction, we have $J(h)=0$.
Hence in this case spin-flip processes do not affect the charge
transfer at all.
This 
contradicts Ref.~[\onlinecite{Dong}], where the current due to
spin-independent tunneling was found to depend on spin-flip rate.
\end{thebibliography}
\end{document}